\documentclass[aps,prl,twocolumn,preprintnumbers,amsmath,amssymb,superscriptaddress]{revtex4-2}
\usepackage{graphicx}
\usepackage[caption=false]{subfig}
\usepackage{epsfig}
\usepackage{dcolumn}
\usepackage{bm}
\usepackage{color}
\usepackage{float}
\usepackage[colorlinks]{hyperref}
\usepackage{verbatim}
\usepackage{algorithm}
\usepackage{algorithmic}

\hypersetup{citecolor = blue}

\def\bra#1{\left\langle#1\right|}
\def\ket#1{\left|#1\right\rangle}

\def\abs#1{\left|#1\right|}

\def\be{\begin{equation}}       \def\ee{\end{equation}}
\def\bea{\begin{eqnarray}}      \def\eea{\end{eqnarray}}
\def\ba{\begin{array}}
\def\ea{\end{array}}
\def\bnum{\begin{enumerate} }
\def\enum{\end{enumerate}}

\def\nn{\nonumber}

\def\=>{\Rightarrow}
\def\>{\rightarrow}

\def\eye2{Fathbb{I}}

\def\Eq#1{Eq.~(\ref{#1})}

\renewcommand{\v}[1]{{\bf #1}}

\renewcommand{\>}{\rangle}

\newcommand{\al}[1]{\begin{align}#1\end{align}}
\newcommand{\eq}[2]{
	\begin{equation}
	#1 \label{#2}
	\end{equation}
}

\newcommand{\re}[1]{\frac{1}{#1}}

\newcommand{\vect}[1]{\boldsymbol{#1}}

\usepackage{listings}
\usepackage{color}
\definecolor{lightgray}{gray}{1}

\begin{document}

\title{Variational Quantum-Neural Hybrid Eigensolver}

\author{Shi-Xin Zhang}
\thanks{The two authors contributed equally to this work.}
\affiliation{Institute for Advanced Study, Tsinghua University, Beijing 100084, China}
\affiliation{Tencent Quantum Laboratory, Tencent, Shenzhen, Guangdong 518057, China}
\author{Zhou-Quan Wan}
\thanks{The two authors contributed equally to this work.}
\affiliation{Institute for Advanced Study, Tsinghua University, Beijing 100084, China}
\affiliation{Tencent Quantum Laboratory, Tencent, Shenzhen, Guangdong 518057, China}
\author{Chee-Kong Lee}
\affiliation{Tencent America, Palo Alto, CA 94306, United States }
\author{Chang-Yu Hsieh}
\email{kimhsieh@tencent.com}
\affiliation{Tencent Quantum Laboratory, Tencent, Shenzhen, Guangdong 518057, China}
\author{Shengyu Zhang}
\affiliation{Tencent Quantum Laboratory, Tencent, Shenzhen, Guangdong 518057, China}
\author{Hong Yao}
\email{yaohong@tsinghua.edu.cn}
\affiliation{Institute for Advanced Study, Tsinghua University, Beijing 100084, China}

\begin{abstract}
	
	The variational quantum eigensolver (VQE) is one of the most representative quantum algorithms in the noisy intermediate-size quantum (NISQ) era, and is generally speculated to deliver one of the first quantum advantages for the ground-state simulations of some non-trivial Hamiltonians. However, short quantum coherence time and limited availability of quantum hardware resources in the NISQ hardware strongly restrain the capacity and expressiveness of VQEs. In this Letter, we introduce the variational quantum-neural hybrid eigensolver (VQNHE) in which the shallow-circuit quantum ansatz can be further enhanced by classical post-processing with neural networks. 
	%The final quantum state output by the VQNHE consistently outperforms that of the VQE when the same amount of quantum resources is allocated for both algorithms in simulating ground-state energy for quantum spins and molecules. 
	We show that VQNHE consistently and significantly outperforms VQE in simulating ground-state energies of quantum spins and molecules given the same amount of quantum resources. 
	More importantly, we demonstrate that for arbitrary post-processing neural functions, VQNHE only incurs an polynomial overhead of processing time and represents the first scalable method to \emph{exponentially} accelerate VQE with non-unitary post-processing that can be efficiently implemented in the NISQ era.

\end{abstract}

\date{\today}
\maketitle

{\bf Introduction:}
Quantum computation was firstly envisioned by Feynman as a natural approach to efficiently simulate quantum systems \cite{Feynman1982}. Equipped with error resilient logical qubits in the fault-tolerant quantum computation regime \cite{Nielsen2010}, we can efficiently and arbitrarily precisely prepare the ground state of any given Hamiltonian by combining adiabatic evolution with quantum phase estimation \cite{Whitfield2011, Georgescu2014}. However, this fault-tolerant strategy requires coherently operating an excessively large number of high-quality physical qubits with high precision that is beyond the realm of current technology \cite{Wecker2014}. Instead, in the NISQ era \cite{Preskill2018}, practical quantum computation is enabled by the hybrid quantum-classical scheme that dramatically alleviates the quantum hardware resources requirement to accomplish non-trivial computational tasks  \cite{Endo2020, Cerezo2020b,Bharti2021}. In such a scheme,  the ground-state problem for a Hamiltonian, $\hat{H}$, is solved by preparing a quantum state in a parameterized quantum circuit (PQC) $U(\vect{\theta})$ as $\vert \psi_\theta\rangle = U(\vect{\theta}) \vert{0}\rangle$. The parameters $\vect{\theta}$ are optimized to minimize $\bra{\psi_\theta}\hat{H}\ket{\psi_\theta}$, and can be trained by a classical optimizer. This hybrid variational approach, usually called VQE, has been successfully applied to a wide range of molecular and quantum spin systems \cite{Peruzzo2014, OMalley2016, McClean2016, Hempel2018, Liu2019b, Cao2019, McArdle2020, Bauer2020},  as well as tasks of excited state search \cite{Higgott2019,Nakanishi2018} and dynamical simulations \cite{Li2017b, Yuan2019, McArdle2019,Cirstoiu2020,  Lin2020, Endo2020a, Benedetti2020, Lee2020}. Indeed, VQE is regarded as one of the most promising routes toward practical quantum advantage \cite{Arute2019,Zhong2020} in the NISQ era.

To fully exploit the potential quantum advantage with VQE, we should design a circuit ansatz having a strong capacity to capture quantum entanglements and correlations possibly present in the target quantum state. Two main categories of circuit ansatz have been proposed for VQE: physics inspired ansatz and hardware efficient ansatz. For physics inspired ansatz, well-established quantum chemistry methods are adapted to the quantum computing context. For example, the probably most famous VQE ansatz, unitary coupled cluster (UCC) ansatz \cite{Taube2006, Peruzzo2014, Wecker2015a, OMalley2016} is inspired by the coupled cluster method, a post Hartree-Fock approach. Though the optimized state from UCC ansatz can in principle give high accuracy when compared to the exact ground state, it requires a very \emph{deep} circuit to implement for the following two reasons: (i) one needs to Trotterize the exponential operator, and (ii) the limited qubit connectivity in many chips, especially the ones with superconducting qubits, introduces a substantial depth overhead. Note that the circuit depth is a crucial measure in the NISQ era in order to accommodate the relatively short coherence time of qubits. To partially address the second issue,  hardware efficient ansatz has been proposed \cite{Kandala2017}. The philosophy is to generate a quantum state by building a PQC with layers of native quantum gates available on a NISQ device and conforms to the hardware's connectivity. While being easily implementable on current quantum hardware, hardware efficient ansatz has been reported to give inferior performance and accuracy \cite{McCaskey2019}. Therefore, in the NISQ era, it is highly desired to devise novel approaches to substantially enhance the performance of VQE while keeping the consumption of quantum resources such as circuit depth and number of quantum gates as low as possible.

\begin{figure*}[t]\centering
	\includegraphics[width=0.95\textwidth]{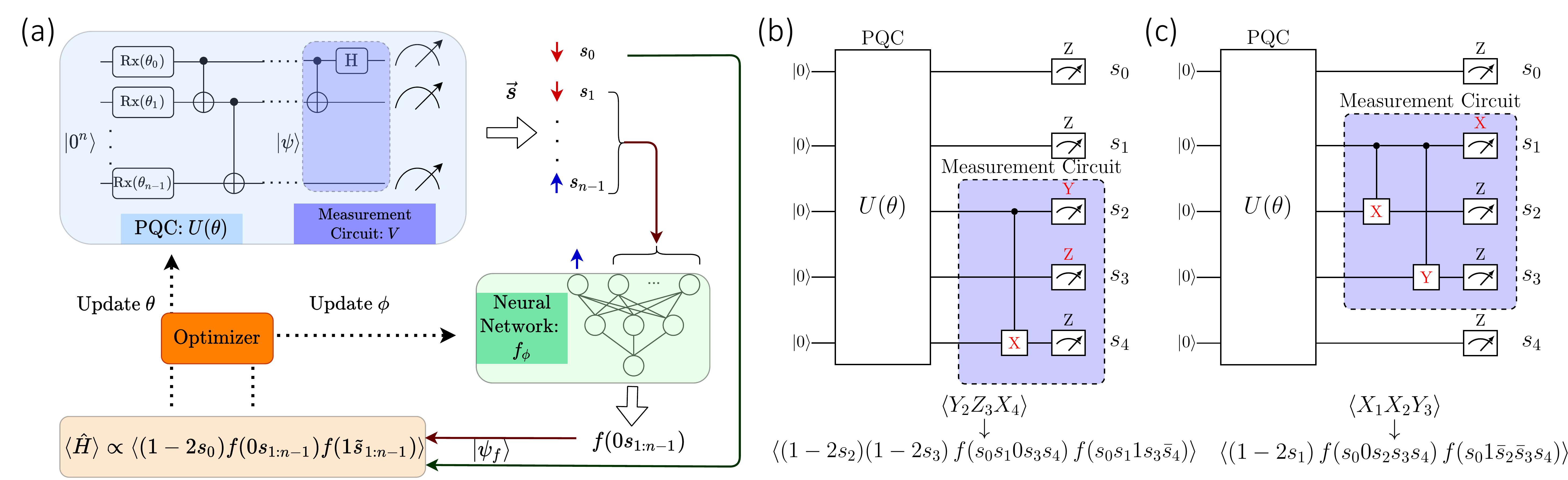}
	\caption{(a) Schematic workflow for VQNHE. The output state $\vert \psi\rangle=U(\vect{\theta})\vert 0\rangle$ is attached with a small measurement circuit $V$. Measurement on computational (Z) basis is conducted on $VU(\vect{\theta})\vert 0\rangle$ to collect shots of bitstring result as $\vec{s}$. With the zeroth qubit as the star qubit (see main text for details), $0s_{1:n-1}$ and $1\widetilde{s}_{1:n-1}$ are fed into the classical neural network $f$ with trainable weights $\vect{\phi}$. The expectation of $\hat{H}$ can then be estimated according to \Eq{eq:main}. Finally parameters in both PQC and neural network are optimized with gradient based optimizer from the expectation result $\langle \hat{H}\rangle$. (b) Measurement protocol for $\hat{H}=Y_2Z_3X_4$.  The star qubit corresponds to $s_2$ and CX gate is applied to qubit 4 since $X_4$ is in $\hat{H}$. Meanwhile, CZ gate is omitted in the hardware level and its effect is counted by the prefactor $s_3$ in the expression. (c) Measurement protocol for $\hat{H}=X_1X_2Y_3$. The star qubit corresponds to $s_1$. CX and CY gates are applied on qubit 2 and 3 respectively, since there are $X_2$ and $Y_3$ in $\hat{H}$.}
	\label{fig:workflow}
\end{figure*}

One promising approach is to combine modern neural networks with quantum circuits in the hybrid quantum-classical paradigm.  A few works have attempted to jointly train a classical neural network with a quantum computing module in order to boost the performance of various tasks \cite{Liu2018d, Liu2019, Verdon2019, Hsieh2019, Benedetti2021}. In particular, the potential gain of introducing classical post-processing to a hybrid algorithm has been actively discussed in the field of quantum machine learning  \cite{Li2020b, Zhang2021}. Different from classification and regression, the task for VQE is to generate a quantum state rather than inferring a label or scalar. Since the quantum nature of the desired output is much harder to be embedded into a classical post-processing framework, very few works have explored this possibility. In Ref. \cite{Mazzola2019}, the authors applied the so-called Jastrow factor \cite{Jastrow1955} $\mathcal{\hat{P}}(\vect{\phi}) = \exp({\sum_{kl}\phi_{kl}Z_kZ_l})$ to the output state $\ket{\psi_\theta}$ of a quantum circuit, yielding the final target state $\ket{\psi_f}= \mathcal{\hat{P}(\vect{\phi})}\ket{\psi_\theta}$. However, existing proposals to supplement standard VQE with $\mathcal{\hat{P}(\vect{\phi})}$ suffer from two main drawbacks. Firstly, though the Jastrow factor is known for capturing quantum correlations in variational Monte Carlo (VMC) \cite{Zen2014, Genovese2019}, it is not the most general form of post-processing and thus the expressive power of such setup is quite limited. More importantly, $\mathcal{\hat{P}}(\vect{\phi})$ cannot be straightforwardly implemented on a quantum computer. Existing methods proposed to evaluate $\bra{\psi_f}\hat{H}\ket{\psi_f}$ require an exponential amount of times or resources to achieve the same measurement accuracy as the standard VQE \cite{Mazzola2019}. %approach the same accuracy as the original VQE. 
Specifically, in the entangled copies method, the probabilities of all bitstring measurements need to be reconstructed, which requires an exponential number of measurements. In the transformed Hamiltonian approach \cite{Benfenati2021}, the extra Jastrow operator is absorbed into the original Hamiltonian $\hat{H}$, and one needs to evaluate $\bra{\psi}\left(\mathcal{\hat{P}}\hat{H}\mathcal{\hat{P}}\right)\ket{\psi}$. Since there is an exponential number of Pauli strings in the Taylor expansion of $\mathcal{\hat{P}}$, one has to evaluate exponential numbers of Pauli strings for the transformed Hamiltonian.

In this Letter, we introduce the variational quantum-neural hybrid eigensolver which falls into the paradigm of variational quantum algorithms enhanced by classical post-processing. Our approach successfully addresses both challenges encountered in the earlier attempts to combine VQE with classical post-processing: (i) VQNHE possesses much greater expressive power as the post-processing can be modeled by any modern neural networks; (ii) VQNHE utilizes the same amount of quantum resource as the original VQE while the classical overhead is provably polynomial in the output range of the neural function and constant in terms of problem size.
We emphasize that the rigorously proven polynomial efficiency of VQNHE is highly nontrivial as the nonunitary post-processing overhead in this scenario is often thought to be intrinsically exponential. Therefore, our approach presents the first scalable method to \emph{exponentially} accelerate VQE with nonunitary post-processing.

{\bf Setup and Method:}
The schematic workflow of VQNHE is shown in Fig.~\ref{fig:workflow}(a).
Suppose the output state from the PQC $U(\vect{\theta})$ is $\ket{\psi}=U(\vect{\theta})\ket{0}$. We propose the following nonunitary post-processing operator:
\eq{\hat{f}=\sum_{s\in\{0,1\}^n} f_\phi(s)\ket{s}\bra{s},}{eq:state-ansatz}
where $f_\phi(s)$ is a parameterized function of a bitstring $s$.  Then the (unnormalized) target output state from VQNHE is $\ket{\psi_f}=\hat{f}\ket{\psi}$. The aim is to minimize the energy expectation
\eq{\langle\hat{H}\rangle_f =\frac{\bra{\psi_f}\hat{H}\ket{\psi_f}}{\bra{\psi_f}\psi_f\rangle}}{eq:vqnheenergy}
by tuning variational parameters $\vect{\theta}$ in the PQC $U$ and $\vect{\phi}$ in the neural network $f$.
When $\hat f$ is applied to $\ket{\psi}$, it adjusts $\psi_s$, the quantum amplitude of $\ket{\psi}$ in the computational basis $s$. For the target ground state $\ket{\psi_0}$, as long as $\{s: \psi_{0s} \ne 0\} \subseteq \{s: \psi_s \ne 0\}$, there always exists an $\hat f$ such that $\hat f \ket{\psi} = \ket{\psi_0}$. In our setup, $f_\phi$ is implemented by a neural network with weights $\vect{\phi}$ which can be learned in training. 
%The function can be taken as any neural function with weights $\vect{\phi}$, namely, we apply $f(s)$ to adjust the quantum amplitude associated with the computational basis $s$. 
For example, Jastrow factor can be regarded as a special case of \Eq{eq:state-ansatz} as $f(s)=\exp({-\sum_{ij} \phi_{ij} (1-2s_i)(1-2s_j)})$.

The key to enabling the above workflow is to efficiently evaluate \Eq{eq:vqnheenergy}.Since $\hat H$ can be decomposed to a summation of Pauli strings, it suffices to to compute the expectation for each Pauli string and then add them up. For this reason, we will assume without loss of generality that $\hat H$ is a Pauli string in the following discussions.
%Suppose $\hat{H}$ is an arbitrary Pauli string without loss of generality (because $\hat{H}$ can be decomposed as a sum of several Pauli strings in the general case, and the analysis still applies). 
As we mentioned in the introduction, na\"ive implementations such as the transformed Hamiltonian approach generally lead to an exponential number of terms to measure which is not practical for any moderate-sized problem.

Firstly, the denominator of \Eq{eq:vqnheenergy} is easy to estimate from measurements as 
\eq{\langle \psi_f\vert \psi_f\rangle=\sum_{s\in\{0,1\}^n} \vert\psi_s\vert ^2\vert f(s)\vert^2.}{eq:den}
The measurement protocol for \Eq{eq:den}  is straightforward: we simply measure the PQC $U$ in computational (Z) basis for multiple shots for $s$, and compute the expectation of $\vert f(s)\vert^2$.
If the Paul string $\hat{H}$ only contains $I$ and $Z$ operators, since $\langle s\vert \hat{H}\vert s'\rangle = H_s\delta_{ss'}$, the estimation for the numerator is also easy:
\eq{\langle \psi_f\vert \hat{H}\vert \psi_f\rangle = \sum_{s\in\{0,1\}^n}\vert \psi_s\vert^2 \vert f(s)\vert^2 H_s.}{eq:n-z}

The key advantage of VQNHE is its efficient scheme to evaluate  $\langle \psi_f\vert \hat{H}\vert\psi_f\rangle$ when $\hat{H}$ contains $X$ or $Y$ operators.  In this case, we label one of the qubits in the Pauli string with X or Y operator as the star qubit, and we rearrange the star qubit as the zero-th qubit in the derivation below.
We further label $\vert \tilde{s}\rangle$ as the bitstring that satisfies $ \hat{H}\vert s\rangle=S(\tilde{s})\vert \tilde{s}\rangle$, where $S=\pm 1, \pm i$ is the sign factor for such a basis transformation under $\hat{H}$. For example, for $\hat{H}=X_0Y_1Z_2$, $\vert \widetilde{011}\rangle = \vert 101\rangle$ and $S(101)=-i$. Since $\hat{H}^2=1$, all eigenvalues are $\pm 1$ and $S(s)S(\tilde{s})=1$. The matrix form of $\hat{H}$ can thus be expressed as
\eq{
\hat{H}=\sum_{
	\mbox{\tiny$\begin{array}{c}
				s_0=0,\\ s_{1:n-1}\in \{0,1\}^{n-1}
\end{array}$}
} S(s)\vert s\rangle\langle \tilde{s}\vert + S(\tilde{s})\vert \tilde{s}\rangle\langle s\vert .
}{}
Note that the sum is over all bitstrings but with the star qubit fixed as $s_0=0$, and we use the shorthand notation $s\in0s_{1:n-1}$ for simplicity.
The $2^{n}$ eigenvectors of $\hat{H}$ with eigenvalue $\pm 1$ have simple forms $\ket{\pm ,  s_{1:n-1}}$:
\al{
\vert +, s_{1:n-1}\rangle &=\frac{1}{\sqrt{2}}(S(0s_{1:n-1})\vert 0s_{1:n-1}\rangle +\vert 1\widetilde{s_{1:n-1}}\rangle)\\
\vert -, s_{1:n-1}\rangle &=\frac{1}{\sqrt{2}}(S(0s_{1:n-1})\vert 0s_{1:n-1}\rangle -\vert 1\widetilde{s_{1:n-1}}\rangle).
\label{eq:eigen}}

%Reve $$\eqref{eigen}$$, we have:
%$$
%\vert s\rangle =\frac{S(\tilde{s})}{\sqrt{2}}(\vert +, s\rangle + \vert -, s\rangle)\\
%\vert \tilde{s}\rangle =\frac{1}{\sqrt{2}}(\vert +, s\rangle - \vert -, s\rangle)
%$$

We restrict $f$ to real-valued functions for now; and for the general case of complex-valued post-processing $f$, efficient estimation is also possible (see the Supplemental Material for details). Then we obtain:

\al{
&\langle \psi_f\vert \hat{H}\vert \psi_f\rangle \nn\\=& \langle \psi\vert \left(\sum_{s\in0s_{1:n-1} }f(s)f(\tilde{s})S(s)\vert s\rangle\langle \tilde{s}\vert+f(s)f(\tilde{s})S(\tilde{s})\vert \tilde{s}\rangle\langle s\vert\right) \vert\psi\rangle
\nn\\= &\langle \psi\vert \left(\sum_{ s\in0s_{1:n-1} } f(s)f(\tilde{s})(\vert +,s\rangle\langle +,s\vert -\vert -,s\rangle\langle -,s\vert )\right)\vert \psi\rangle \nn\\=& \sum_{ s\in0s_{1:n-1} } \vert\psi_{+,s}\vert^2 f(s)f(\tilde{s}) + \vert\psi_{-,s}\vert^2 (-f(s)f(\tilde{s}))
,}
where $\psi_{\pm,s} = \langle \pm, s_{1:n-1}\vert\psi\rangle$.

To realize a measurement in the eigenbasis of $\hat{H}$, we attach a measurement circuit $V$ after the original PQC $U(\vect{\theta})$ such that the computational basis measurement on the output from $VU(\vect{\theta})\ket{0}$ corresponds to the amplitude for $\vert \pm , s_{1:n-1}\rangle$, where the readout for first (star) qubit  represents the eigenvalue of $\hat{H}$ and the readout for the remaining $n-1$ qubits  stand for $s_{1:n-1}$. Specifically, we require $V^\dagger \vert{s}\rangle\propto \vert \pm , s_{1:n-1}\rangle$ so that $\psi_{\pm, s} = \langle \pm , s_{1:n-1}\vert \psi\rangle = \langle{s}\vert V U(\vect{\theta})\vert 0\rangle$. The problem is now reduced to efficiently building a measurement circuit $V$ which gives $V^\dagger\vert s\rangle\propto (\vert 0s_{1:n-1}\rangle +(1-2s_0)\hat{H}\vert 0s_{1:n-1}\rangle)
$.
We now describe how to build this $V$ circuit:
\begin{enumerate}

		\item For all qubits present in the Pauli string $\hat{H}$ except the star qubit, we apply a control-X/Y/Z gate with the control being the star qubit, and the choice of the control gate is determined by the Pauli operator acting on the corresponding qubit in $\hat{H}$. (Note that control-Z gate application can be omitted and replaced by counting the extra sign $s_i$ in the final expression.)
		\item The star qubit is measured in the $X$ or $Y$ basis determined by the corresponding operator in $\hat{H}$, or equivalently speaking, the star qubit is attached with a single-qubit gate: Hadamard gate in $X$ case, and Rx$=\exp(-\pi/4\; i X)$ rotation gate in $Y$ case, and then measured on the computational (Z) basis.
\end{enumerate}
We explicitly constructed the measurement circuit $V$ for a few representative Pauli string $\hat{H}$, as shown in Fig.~\ref{fig:workflow}(b) and (c).
By appending the aforementioned compact measurement circuit $V$ to $U(\theta)$ and collecting measurement results as bitstring $s$, the expectation value from the quantum-neural hybrid state is given by: 
\eq{
\langle \hat{H}\rangle_{\psi_f} = \frac{\langle{(1-2s_0) f(0s_{1:n-1})f(1\widetilde{s_{1:n-1}})) }\rangle_{UV}}{\langle{f(s)^2}\rangle_{U}},
}{eq:main}
where bitstring $s$ in the denominator is drawn from the PQC $U$ and bitsring $s$in the numerator is drawn from the PQC with the measurement circuit $V$ appended.

The extra quantum resources compared to the original PQC for VQE is at most $m-1$ two-qubit gates, where $m$ is the number of $X$ and $Y$ operators in the Pauli string $\hat{H}$. For typical short-range interaction Hamiltonians, we note that $m=O(1)$. Besides, the number of measurement shots required to achieve the same accuracy as VQE is polynomial bounded in the VQNHE setup (see the Supplemental Materials for details of a rigorous proof). Now that we can efficiently evaluate $\langle \hat{H}\rangle_{\psi_f}$, the gradients with respect to the PQC and the neural network can be efficiently obtained via parameter shift \cite{Li2017d, Mitarai2018, Schuld2019a} and backpropagation, respectively, which facilitate gradient-based classical optimizers to update parameters $\vect{\theta}$ and $\vect{\phi}$.

With the presented formalism and protocol, we have demonstrated that VQNHE, the combination of variational quantum eigensolver and classical nonunitary neural post-processing,  gives rise to an \emph{exponential} acceleration compared to previous methods when incorporating non-unitary post-processing into the standard VQE.

{\bf Results:}
In this section, we report the performance of VQNHE on several benchmarks in modeling quantum spins and molecules, including 1D transverse field Ising model (TFIM), 1D Heisenberg model, LiH, H$_6$-hexagon and H$_6$-chain molecule \footnote{The open source implementation of VQNHE can be found at \url{https://github.com/refraction-ray/tensorcircuit}}. (See the Supplemental Material for details on the setup and results for each system.)

First, we present numerical results for quantum spin models: the TFIM defined as $H_{\text{TFIM}}=\sum_{i,i+1}Z_iZ_{i+1}-\sum_i X_i$, and the Heisenberg model defined as $H_{\text{Heisenberg}}=\sum_{i, i+1}(X_iX_{i+1}+Y_iY_{i+1}+Z_iZ_{i+1})$, both imposed with the periodic boundary condition. We apply both VQNHE and VQE to simulate the ground state of these systems with $N=12$ sites. %For the PQC part, we adopt the multi-parameter version of QAOA ansatz \cite{Farhi2014} for TFIM and parameterized swap layered ansatz for Heisenberg model to take SU(2) symmetry of the Heisenberg model into consideration. 
The results for the ground-state energy of these two systems are summarized in Table.~\ref{table:vqecomparison}. Note that VQNHE provides a substantially more accurate estimation (about two orders of magnitude improvement in terms of energy estimation accuracy) of the ground-state energy using the same amount of quantum resources.

\begin{table}[]\centering

%	\begin{tabular}{cccc}
%\hline
%		Model                & VQE                     &VQNHE         & exact           \\
%\hline
%	TFIM & -14.914 ($3*10^{-2}$) &  -15.319 ($2*10^{-4}$)    & -15.3226\\
%	Heisenberg Model   &       -21.393 ($7*10^{-3}$) &        -21.546 ($2*10^{-4}$)            & -21.5496     \\
%	
%\hline
%	\end{tabular}

	\begin{tabular}{ccc}
	\hline
	Model                & TFIM                  &Heisenberg Model          \\
	\hline
	VQE &   -14.914 ($3*10^{-2}$) &     -21.393 ($7*10^{-3}$)              \\
	VQNHE &  -15.319 ($2*10^{-4}$) &  -21.546 ($2*10^{-4}$)           \\
	exact &-15.3226&-21.5496\\
	\hline
\end{tabular}

	\caption{The ground state energies obtained from both VQE and VQNHE with the same PQC structure. for 1D TFIM and Heisenberg model with $N=12$ sites. Relative errors compared to the exact ground state are included. For both models, the energy obtained from VQNHE is much closer to the exact ground state
	energy than the one obtained from VQE  }
	\label{table:vqecomparison}
\end{table}

\begin{figure}[t]\centering
	\includegraphics[width=0.47\textwidth]{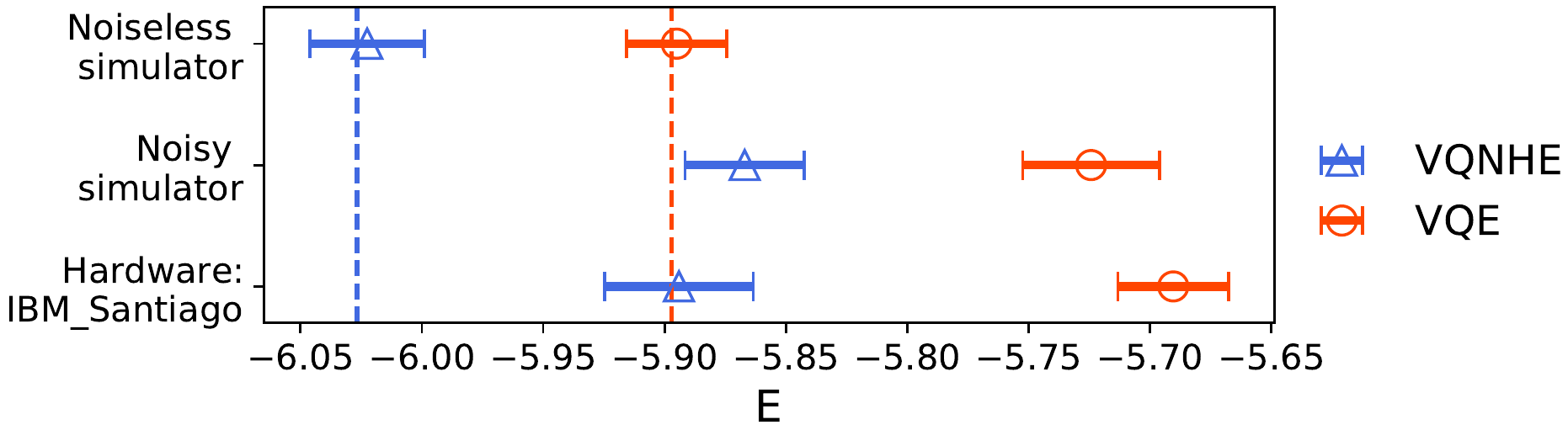}
	\caption{Optimized energies on the 5-site TFIM model with open boundary condition using either VQNHE (blue) and VQE (red) approaches.  Measurement-based results in noiseless, noisy simulators, and real quantum hardware are shown with the error bar. The vertical red and blue lines are the ideal optimized energy values from VQNHE and VQE, when no quantum noise or measurement uncertainty exists. The blue line also coincides with the exact ground state energy for the model since the ideal VQNHE result only has a relative error in the order of $10^{-12}$. }
	\label{fig:accbar}
\end{figure}

We further obtain the optimized energies from the VQE and VQNHE algorithms simulated on IBM quantum hardware and noisy simulators, as shown in Fig.~\ref{fig:accbar}. The target model is the 5-site TFIM model with open boundary condition. The results demonstrate that the VQNHE works well in the presence of quantum noise and measurement uncertainty on the real quantum hardware, namely, VQNHE exhibits a strong noise resilience because of the further optimization done with the post-processing module.
%VQNHE can also reduce the effect of quantum noise via further optimization of the post-processing module.

We now turn to the calculation of the energy dissociation curve for LiH, another common benchmark for VQE. The task is to calculate the ground-state energy of LiH at different bond distances. To obtain the Hamiltonian of LiH in terms of qubit operators, we firstly preprocess LiH using procedures including overlap integral calculation and qubit encoding for a fermionic Hamiltonian. These procedures are done using Psi4 \cite{Turney2012} and OpenFermion \cite{McClean2017}, respectively. By using symmetry enforced binary code \cite{Steudtner2017} on the complete active space of Hartree-Fock molecular orbitals, we obtain a 4-qubit Hamiltonian composed of 100 Pauli string terms.

\begin{figure}[t]\centering
	\includegraphics[width=0.36\textwidth]{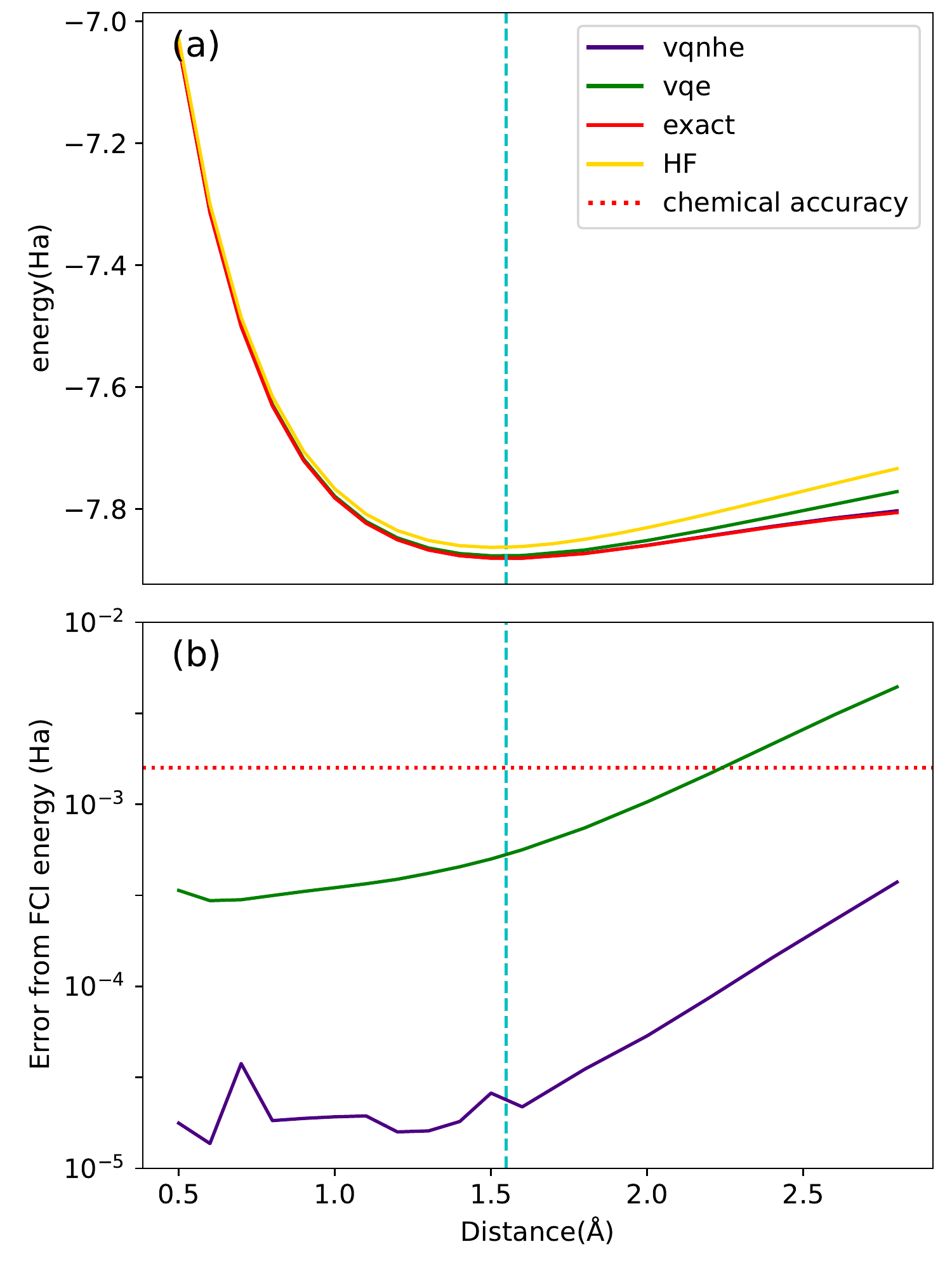}
	\caption{LiH dissociation curve. (a) VQNHE optimized energy (purple), VQE converged energy (green), exact energy from FCI (red), Hartree-Fock energy (yellow) obtained at different bond distances with STO-3G basis set and symmetry binary encoding within complete active space. (b) Comparison of corresponding energy errors for VQNHE and VQE results. The red dash line is the threshold of chemical accuracy. VQNHE energy is always within chemical accuracy in the whole bond distance range. The cyan vertical line is the bond distance with minimum bonding energy and represents LiH molecule at equilibrium configuration.}
	\label{fig:lih}
\end{figure}

We optimize LiH qubit Hamiltonian using VQNHE and VQE with 20 independent runs for each bond distance from 0.5 to 2.8\r{A}, and the best results of each instance are reported in Fig.~\ref{fig:lih}. Both VQNHE and VQE utilize the same $\text{depth-2}$ hardware efficient ansatz. The relative error of the VQNHE result is in the order of $10^{-5}$ 
and this result matches the state-of-the-art result given by RBM-based VMC \cite{Choo2020}. For comparison, vanilla VQE can only achieve a relative error around the order of $10^{-3}$.

Furthermore, we apply VQNHE on the molecular system H$_6$-hexagon and H$_6$-chain. Via symmetry enforced qubit encoding, we can simulate the corresponding system with a 10-qubit PQC and complex-RBM based post-processing module.  The relative errors of optimized energy for both systems are in the order of $10^{-5}$ and $10^{-6}$, respectively. Our VQNHE results are not only within chemical accuracy, but actually outperform CCSD method. 

{\bf Discussions:}
The VQNHE presented in this Letter sits at the intersection between VQE and VMC \cite{Hill1965, Ceperley1977,Carleo2017}. It is similar to the VMC setup for complex-valued wavefunction, where two computational graphs are utilized: one is for the amplitude and the other one is for the phase or sign structure. Since the quantum phase is harder to characterize than the amplitude \cite{Westerhout2019,Bukov2020,Park2020}, tensor network ansatzes have been proposed to capture such subtlety as a replacement of the neural network \cite{Liang2020}. Within the VQNHE framework, we can view the PQC $U(\vect{\theta})$ as the part responsible to learn quantum phase, taking a similar role as the tensor networks in the VMC example described above. Since the PQC is quantum by nature, it is expected to better capture quantum entanglement and learn the quantum phase structure of the target state more efficiently.  Besides, sampling from the PQC is highly efficient as it can draw independent samples each time without a high rejection ratio in traditional Metropolis-Hasting sampling strategy. 
Despite the similarity between VQNHE and VMC,  VQNHE cannot be efficiently implemented within a VMC framework, since $\langle s\vert \psi\rangle$ in the denominator of $E_{\text{loc}}$ in VMC costs exponential numbers of measurement shots to accurately estimate $E_{\text{loc}}$.  This observation further bolsters the significance of our efficient protocol for VQNHE.
In summary, the VQNHE approach can be either referred to as the neural-network enhanced VQE or as the quantum-computing assisted VMC; it actually combines the advantages of both.

One of the promising future directions is to combine VQNHE and quantum architecture search  \cite{Zhang2020b,Li2020,Du2020a, Lu2020,Bilkis2021,Zhang2021,Kuo2021} or adaptive VQE \cite{Grimsley2019, Grimsley2019, Rattew2019, Chivilikhin2020,Yordanov2020, Sim2020,Claudino2020} in which the parameterized circuit ansatz can be iteratively adjusted or grown to improve the overall performance for such a hybrid workflow. Moreover, it is worth investigating theoretically whether VQNHE is more robust against quantum noise than the vanilla VQE, as relevant evidences emerge from hardware experiments in this work.

\textbf{Conclusion:} In this Letter, we propose VQNHE that combines nonunitary post-processings with the PQC to improve upon VQE. VQNHE uses a hybrid representation of quantum states in order to enhance the expressive power with limited quantum hardware resources, and it consistently outperforms VQE in various tasks.  We also outline a feasible protocol to implement VQNHE on real quantum hardware with rigorously proved efficiency. We demonstrate that VQE with arbitrary nonunitary post-processing can be accurately carried out with only polynomial overhead: an \emph{exponential} improvement of efficiency that was deemed unlikely along the lines of prior proposals.  %general belief and previous works.

~\newline
\textbf{Acknowledgements:} This work is supported in part by the NSFC under Grant No. 11825404 (SXZ, ZQW, and HY), the CAS Strategic Priority Research Program under Grant No. XDB28000000 (HY), and  Beijing Municipal Science and Technology Commission under Grant No. Z181100004218001 (HY).

%\bibliographystyle{/Users/shixin/Cloud/refdb/refdatabase/apsreve.bst}
%\bibliography{/Users/shixin/Cloud/refdb/refdatabase/lib2/library.bib}
%merlin.mbs apsrev4-1.bst 2010-07-25 4.21a (PWD, AO, DPC) hacked
%Control: key (0)
%Control: author (72) initials jnrlst
%Control: editor formatted (1) identically to author
%Control: production of article title (-1) disabled
%Control: page (0) single
%Control: year (1) truncated
%Control: production of eprint (0) enabled
%

\newpage

\begin{widetext}
	\section*{Supplemental Materials}
	\renewcommand{\theequation}{S\arabic{equation}}
	\setcounter{equation}{0}
	\renewcommand{\thefigure}{S\arabic{figure}}
	\setcounter{figure}{0}
	
	\subsection{Measurement efficiency for VQNHE}
	As shown in the main text, the overhead of quantum hardware resources for VQNHE  is at most $m-1$ two-qubit gates where $m\leq n$ is the number of X and Y operators present in the Pauli string $\hat{H}$. Furthermore, our proposed protocol is also unbiased without any systematic deviations from the ground truth. Therefore, the only concern left is: how many measurement shots are required to estimate the energy expectation at a given precision? If the number of shots required is only polynomial times more than the original VQE measurements, then we may conclude that VQNHE is efficient and exponentially faster than previous proposals attempting to incorporate nonunitary post-processing with VQE.  
	
	For the vanilla VQE, assuming that there is probability $p$ to measure $\hat{H}$ with $+1$ result, then the standard deviation for the estimation after $N$ measurement shots is:
	\eq{\delta \langle \hat{H}\rangle = \frac{2\sqrt{p(1-p)}}{\sqrt{N}},}{}
	which is due to the deviation formula for the binomial distribution. Therefore, to achieve an accuracy of $1-\varepsilon$, we have to measure the system at least $N=\frac{4p(1-p)}{\varepsilon^2}$ times. The hardest case is when $\langle \hat{H}\rangle=0$ with $p=1/2$, and the required number of measurement shots is in the order of $\frac{1}{\varepsilon^2}$.
	
	Next, we turn to VQNHE, the expectation value is computed as a fraction of two quantities, $\langle \hat{H}\rangle = n/d$, where $d=E_{p(s)}(f^2(s))$ and $n=E_{q(s)}((1-2s_0)f(0s_{1:n-1})f(1\widetilde{s_{1:n-1}}))$. $p(s)$ and $q(s)$ are two distributions determined by the output state of two different PQCs $U$ and $VU$. Considering the standard deviation, we have:
	\eq{\delta \langle \hat{H}\rangle=\delta (\frac{n}{d})< \vert \frac{\delta n}{d\sqrt{N}} \vert+ \vert \frac{\delta d \,n}{d^2\sqrt{N}}\vert, }{}
	where $\delta n$ and $\delta d$ are the standard deviation of the random variable in the denominator and numerator. Recall $\langle \hat{H}\rangle = n/d<1$ and suppose the neural function f is restricted in the range $[1/r, r]$, then we have $\frac{1}{d}\leq r^2$, $\delta d< (r^2-1/r^2)/2$, $\delta n< 2r^2 \sqrt{p(1-p)}$ and
	\eq{\delta  \langle \hat{H}\rangle< \frac{1}{\sqrt{N}} (\frac{\delta d+\delta n}{d})<\frac{3r^4}{2\sqrt{N}}.}{}
	Therefore, the required number of shots to attain the same accuracy $\varepsilon$ is at most $N=\frac{9}{4}r^8/\varepsilon^2$.
	
	We can further tight the bound by more thorough analysis. In terms of the first term:
	\eq{\left(\frac{\delta d}{d}\right)^2 =\frac{ \langle  f^4(s)\rangle_s}{\langle f^2(s) \rangle_s^2}-1, }{eq:ddoverd}
	where the expectation is defined by bitstring $s$ collected from the PQC $U$.
	The upper bound of \Eq{eq:ddoverd} is reached when $f(s)$ is a two-value function always giving $r$ or $1/r$. This is because, otherwise, pairs of $f(s_1)$ and $f(s_2)$ of values in $(1/r, r)$ can be pushed to two sides in a way that leaves $E(f^2(s))$ unchanged until one of the two $f$ values is now in the boundary. This ``push'' procedure keeps the denominator of \Eq{eq:ddoverd} unchanged while monotonically increases the numerator of \Eq{eq:ddoverd}. 
	Therefore, we have:
	\eq{\left(\frac{\delta d}{d}\right)^2\leq\max_p \frac{p r^4+(1-p)1/r^4}{(p r^2+(1-p)/r^2)^2}-1 =\frac{r^4}{4}+\frac{1}{4 r^4}+\frac{1}{2}-1,}{}
	where $p$ is the total probability to measure $\{s\vert f(s)=r\}$. Hence, 
	\eq{\frac{\delta d}{d}\leq \frac{1}{2}(r^2-\frac{1}{r^2}).}{}
	
	We now focus on the second term $\delta n/d$. We define the probability $p_{s_{1:n-1}} = \vert \langle 0s_{1:n-1}\vert \psi\rangle\vert ^2 +\vert \langle 1\widetilde{s_{1:n-1}}\vert \psi\rangle\vert ^2$, where $\ket{\psi}$ is the wavefunction from the PQC $U$. In the numerator, we require the probability amplitude of $\psi_{\pm, s_{1:n-1}}=\langle \pm, s_{1:n-1}\vert \psi\rangle$ which is just $\re{\sqrt{2}}(\langle 0s_{1:n-1}\vert \psi\rangle \pm S(1\widetilde{s_{1:n-1}}) \langle 1\widetilde{s_{1:n-1}}\vert \psi\rangle)$. We define the normalized probability amplitude as:
	\eq{\phi_{\pm, s_{1:n-1}}=\frac{1}{\sqrt{p_{s_{1:n-1}}}}\psi_{\pm, s_{1:n-1}} =\re{\sqrt{2p_{s_{1:n-1}}}}(\langle 0s_{1:n-1}\vert \psi\rangle \pm S(1\widetilde{s_{1:n-1}}) \langle 1\widetilde{s_{1:n-1}}\vert \psi\rangle), }{}
	with $p_{s_{1:n-1}}^{-1/2}$ as the normalization factor. We then have $\abs{\phi_{+, s_{1:n-1}}}^2+\abs{\phi_{-, s_{1:n-1}}}^2=1$. Namely, to accommodate the probability distribution in the numerator and the denominator, we recast the distribution freedom as a classical part $p_{s_{1:n-1}}$ and a quantum part $\phi_{\pm, s_{1:n-1}}$. These two parts are decoupled and vary independently. We now have:
	\eq{\left(\frac{\delta n}{d}\right)^2 = \frac{\langle ((1-2 s_0) f(0s_{1:n-1})f (1\widetilde{s_{1:n-1}}))^2\rangle_{UV}}{\langle f^2(s) \rangle_U^2}-\langle \hat{H}\rangle^2\leq \frac{\langle (f(0s_{1:n-1})f (1\widetilde{s_{1:n-1}}))^2\rangle_{UV}}{\langle f^2(s) \rangle_U^2},}{}
	where we have utilized the fact that $\langle \hat{H}\rangle ^2 \in [0,1]$. The expectation denoted by $\langle \cdot\rangle_{UV}$ and $\langle \cdot\rangle_{U}$ is computed by the measured bitstrings from the PQC $U$ with and without the measurement circuit $V$. Note the probability to get $s$ from the PQC $U$ is $\frac{p_{s_{1:n-1}}}{2}\vert \phi_{+, s_{1:n-1}} +(1-2s_0)\phi_{-, s_{1:n-1}}\vert^2 $:
	\al{\left(\frac{\delta n}{d}\right)^2&\leq \frac{\sum_{0s_{1:n-1}} p_{s_{1:n-1}} f^2(0s_{1:n-1})f^2(1\widetilde{s_{1:n-1}} ) }{ 1/4(\sum_{0s_{1:n-1}}   p_{s_{1:n-1}} ( f^2(0s_{1:n-1}) \vert \phi_++\phi_-\vert ^2 +f^2(1\widetilde{s_{1:n-1}} ) \vert \phi_+-\phi_-\vert^2 ))^2 } \nn\\
& = 4\frac{ \sum_{0s_{1:n-1}} p_{s_{1:n-1}} f^2(0s_{1:n-1})f^2(1\widetilde{s_{1:n-1}} ) }{
(\sum_{0s_{1:n-1}}   p_{s_{1:n-1}} ( f^2(0s_{1:n-1})+f^2(1\widetilde{s_{1:n-1}}  )  +2(f^2(0s_{1:n-1})-f^2(1\widetilde{s_{1:n-1}} )) Re(\phi_+^*\phi_-) ) )^2
}.}

Since $\phi_\pm$ is only in the denominator,  we let $2Re(\phi_+^*\phi_-)=-\text{sign}(f^2(0s)-f^2(1\tilde{s}))$ to make the denominator as small as possible (we use $0s$ and $1\tilde{s}$ as the shortcut for $0s_{1:n-1}$ and $1\widetilde{s_{1:n-1}} $,  respectively). We now have:
\al{\left(\frac{\delta n}{d}\right)^2 &\leq \frac{
\sum_{0s} p_{s} f^2(0s)f^2(1\tilde{s} ) 	
}{ (\sum_{0s} p_{s} \min(f(0s), f(1\tilde{s}))^2)^2}\nn\\
&= \frac{
	\sum_{0s} p_{s} f^2(0s)f^2(1\tilde{s} ) 	
}{ \sum_{0s} p_{s} \min(f(0s), f(1\tilde{s}))^2} \frac{1}{ \sum_{0s} p_{s} \min(f(0s), f(1\tilde{s}))^2}\nn\\
&\leq r^2 \frac{
	\sum_{0s} p_{s} f^2(0s)f^2(1\tilde{s} ) 	
}{ \sum_{0s} p_{s} \min(f(0s), f(1\tilde{s}))^2}.}
Without loss of generality, we assume $f(1\tilde{s})>f(0s)$ for each $s_{1:n-1}$, then we let $f(1\tilde{s})=r$ to maximize the above formula:
\eq{\left(\frac{\delta n}{d}\right)^2 \leq r^4\frac{\sum_{0s}p_{s} f^2(0s)}{\sum_{0s}p_{s} f^2(0s)}=r^4.}{}
	
	To sum up, we have:
	\eq{\delta  \langle \hat{H}\rangle< \frac{1}{\sqrt{N}} (\frac{\delta d+\delta n}{d})\leq\frac{1}{\sqrt{N}}(\frac{r^2}{2}-\frac{1}{2r^2}+r^2)\leq\frac{1}{\sqrt{N}}\frac{3r^2}{2}.}{}
	Therefore, the required number of measurement shots to reach the accuracy $1-\varepsilon$ is at most $N=\frac{9 r^4}{4}\frac{1}{\varepsilon^2}$.
	
	This bound is tight in terms of the scaling with $r$, as one can devise $f$ and the PQC $U$ such that the bound saturates. The extra overhead for VQNHE (with respect to VQE) is only polynomial in terms of the cutoff $r$ and independent of the system size $N$. Therefore,  VQNHE is highly efficient compared to previous proposals that require an exponential amount of resources. In actual experiments, as long as the output of $f$ is restricted to a reasonable range, the overhead should be much smaller than the one indicated by the upper bound here. This is because, in real problems, the distribution $p$ based on the wavefunction and the distribution of $f$ value over the bitstring space is not that extreme.

	\subsection{VQNHE formalism with complex neural function}
	In this section, we show an efficient measurement protocol for complex-valued neural post-processing as a complement to the result in the main text. 
We have:
	\al{&\langle \psi_f\vert \hat{H}\vert \psi_f\rangle \nn\\=& \langle \psi\vert \left(\sum_{s\in0s_{1:n-1}}f^*(s)f(\tilde{s})S(s)\vert s\rangle\langle \tilde{s}\vert+f(s)f^*(\tilde{s})S(\tilde{s})\vert \tilde{s}\rangle\langle s\vert\right) \vert\psi\rangle
		\nn\\= &\bra{\psi}  \sum_{s\in0s_{1:n-1}} R(s) (S(s)\vert s\rangle\langle \tilde{s}\vert+S(\tilde{s})\vert \tilde{s}\rangle\langle s\vert) +i I(s) (S(s)\vert s\rangle\langle \tilde{s}\vert-S(\tilde{s})\vert \tilde{s}\rangle\langle s\vert)\ket{\psi}, \label{eq:complex}}
	where $R(s)=Re(f^*(s)f(\tilde{s}))$ and $I(s) = Im(f^*(s)f(\tilde{s}))$. For the first term in the last line of \Eq{eq:complex},  one may simply use the measurement and estimation protocol given in the main text with a simple replacement: substituting $f(0s_{1:n-1})f(1\widetilde{s_{1:n-1}})$ with $Re(f(0s_{1:n-1})f(1\widetilde{s_{1:n-1}}))$.
	
	We now focus on the second term in the last line of \Eq{eq:complex},
	\al{&\bra{\psi}\sum_{s\in0s_{1:n-1}}iI(s) (S(s)\vert s\rangle\langle \tilde{s}\vert-S(\tilde{s})\vert \tilde{s}\rangle\langle s\vert)\ket{\psi}\nn\\ =&
\bra{\psi} \sum_{s\in0s_{1:n-1}}I(s) [\ket{+, s}'\bra{+,s}'-\ket{-,s}'\bra{-, s}'] \ket{\psi}	\nn\\ =&
\sum_{s\in0s_{1:n-1}} \vert \psi_{+,s}'\vert ^2 I(s) +\vert \psi_{-,s}' \vert^2(-I(s))),
}
where $\psi_{\pm,s}' = \langle \pm ,s\vert' \psi\rangle$ are the amplitudes associated with a new basis $\vert \pm , s_{1:n-1}\rangle'$:
\al{
\vert +, s_{1:n-1}\rangle'=&\frac{1}{\sqrt{2}}(-iS(s)\ket{0s_{1:n-1}}-\ket{1\widetilde{s_{1:n-1}}}) \nn \\
\vert -, s_{1:n-1}\rangle'=&\frac{1}{\sqrt{2}}(-iS(s)\ket{0s_{1:n-1}}+\ket{1\widetilde{s_{1:n-1}}}).
}

Therefore, to measure the contribution from the imaginary part of $f$, we need another measurement circuit $V'$ such that $(V')^\dagger \vert s\rangle \propto \vert \pm, s_{1:n-1}\rangle' \propto \frac{1}{\sqrt{2}}(-i\ket{0s_{1:n-1}} - (1-2s_0) \hat{H}\ket{0{s_{1:n-1}}}) $. Clearly, $V'$ can be constructed with minor modifications of our proposed method to build $V$ in the main text. Firstly, we still have to locate the star qubit and apply control-X/Y/Z gates on all other qubits involved in the Pauli string $\hat{H}$. The only difference is that we measure in the X(-Y) basis for the star qubit when the operator is Y(X) in the $\hat{H}$. In other words, we apply Hadamard gate on the star qubit if Y operator is present in $\hat{H}$, and apply $\text{Rx}= \exp(\pi/4\; i X)$ on the star qubit if X operator is present. In short, the protocol to estimate the expectation value from quantum-neural hybrid state costs only twice the processing time as the case for the real-valued $f$.
	
	\subsection{Technical details for the simulation of quantum spin systems}
	
	1D TFIM and Heisenberg model with 12 sites and periodic boundary conditions are evaluated by both VQE and VQNHE. The quantum ansatz used in the TFIM case is a multiple parameter version of the QAOA ansatz with alternating ZZ and X layers:
	\eq{{U}(\vect{\theta}) = \prod_{p=1}^2 \left(\prod_{i=1}^{12} e^{i\theta_{ip,x} X_i}\prod_{i=1}^{12} e^{i\theta_{ip,zz}Z_iZ_{(i+1)\%12}} \right)\prod_{i=1}^{12}H_i,}{}
	where $H_i$ is the Hadamard gate on the i-th qubit.
	
	The PQC we utilized in the main text for isotropic Heisenberg model is built with layers of parametrized swap gates with pairs of Bell states as the input. Specifically, we have:
	\eq{{U(\vect{\theta})}\vert 0^{12}\rangle = \prod_{p=1}^2 \left( \prod_{i=1}^{12} e^{i\theta_{ip} \text{SWAP}_{i, i+1}} \right) \prod^6 (\frac{\vert 01 \rangle-\vert 10\rangle}{\sqrt{2}}),}{}
	where $\text{SWAP}=\frac{1}{2}(X_1X_2+Y_1Y_2+Z_1Z_2+I_1I_2)$. This ansatz conserves the SU(2) symmetry for the isotropic Heisenberg model and thus keep the variational state in the same symmetry sector as the ground state. 
	
	The classical neural model we utilized is a real-valued fully connected neural network with two hidden layers of  width 24 and 12 with ReLU activation (we attach one more 24-unit hidden layer for the Heisenberg model). The final scalar output is activated via $e^{\phi_0\tanh(*)}$, where $\phi_0$ is a trainable weight that regulates the output range of $f$ and, in turn, controls the magnitude of fluctuations for VQNHE estimation. By restricting the range or the maximum value of $f$, we can keep the measurement overhead for VQNHE estimation as low as possible while witnessing just a slightly worse performance. This is a classic trade-off. Nonetheless, we report the result for a Heisenberg model with $\phi_0<1$ imposed and $f\in [1/e, e]$ is guaranteed. The required number of measurement shots for the given accuracy is only three times more than that for the original VQE. And the approximation ratio in the result is still very good: VQNHE gives $-21.53$ (0.1\% in relative error) for the ground-state energy of the Heisenberg model, which is still much better than the vanilla VQE results (given by an ansatz with 2-layer parameterized swap gates).
	
	For the same Heisenberg model, we also report VQE and VQNHE results given by the hardware efficient ansatz ([Rx, Rz, CNOT]*2). The na\"ive VQE gives the converged energy $-19.26$ while VQNHE gives the converged energy $-21.30$. The performance boost compared to VQE is more significant when we do not explicitly impose symmetry in the circuit architecture.
	
	It is also interesting to observe how the PQC incorporates the phase structure of Heisenberg model's ground state function. Since without transformation of $YY$ coupling, the ground state of Heisenberg model shows a sign structure. The relevant information learned by the PQC in a VQNHE setup is shown in Fig.~\ref{fig:sign}.
	
	\begin{figure}[t]\centering
		\includegraphics[width=0.45\textwidth]{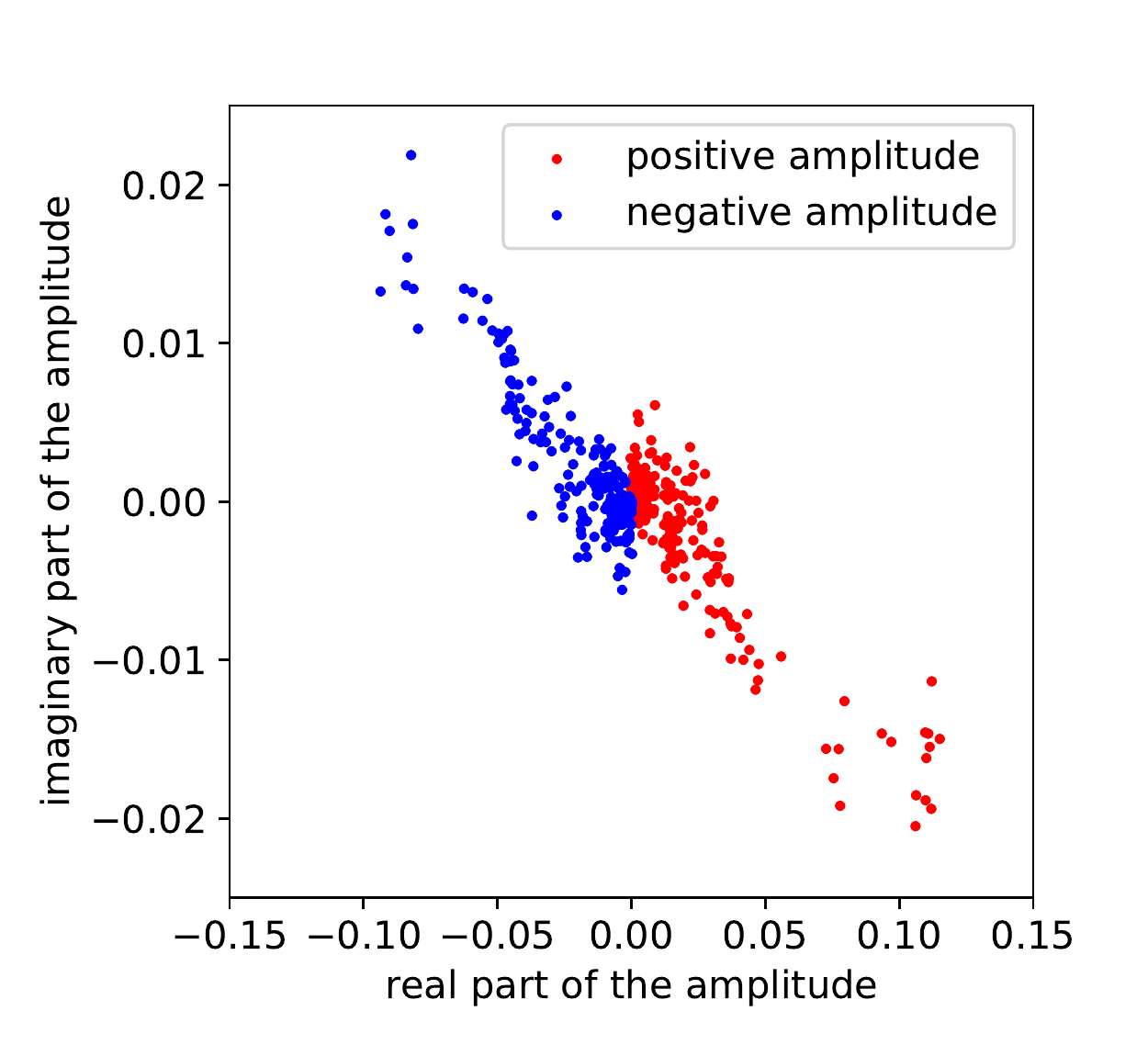}
		\caption{Sign structure learned by the PQC in VQNHE setup. The scatter points represent the amplitude components on the computational basis from the optimized PQC wavefunction: real and imaginary parts are shown. The color on each point stands for the sign of the exact ground state. As we can see, the PQC roughly learns about the sign structure of Heisenberg model's ground state up to a phase. }
		\label{fig:sign}
	\end{figure}
	
	In addition, we comment on the joint training of quantum and classical modules. Firstly, the parameters in the PQC and the neural network should be updated with different optimizers as the learning tend to progress at different scales. We choose Adam optimizer for all simulations in this work. Secondly, we observe that the na\"ive joint training and stage-wise training where the PQC and the neural network get optimized sequentially, are both inferior optimization approaches. Instead, a much better solution is to optimize the PQC first then follows by a joint optimization on both the PQC and the neural network (but the PQC receives a much smaller learning rate in the second stage). An example of the learning curves for both optimizers (PQC and neural network) is displayed in Fig.~\ref{fig:lclih}.

	\subsection{Technical details for quantum hardware experiments}
	We use the 1D 5-site TFIM model with open boundary condition as the test system, since the quantum hardware provided by IBM (specifically we use IBM\_Santiago instance) shows a one dimensional connectivity. The exact ground state energy for this model obtained from exact diagonalization is $-6.02667418$. The PQC we utilized in both VQE and VQNHE experiments are shown in Fig.~\ref{fig:santiago}. In the design of the PQC, the number of two-qubit gates is kept small for high-fidelity results, as the quantum error brought by two-qubit gates is rather large in real hardware. So we only keep one layer of $e^{i\theta Z_iZ_j}$ gates. The classical neural model in the VQNHE setup is a real-valued fully connected neural network with two hidden layers of 10 and 20 units, and the activations for the two layers are ReLU and sigmoid, respectively. The final scalar output is further activated via $e^{\phi_0\tanh (*)}$. 
	
		\begin{figure}[t]\centering
		\includegraphics[width=0.55\textwidth]{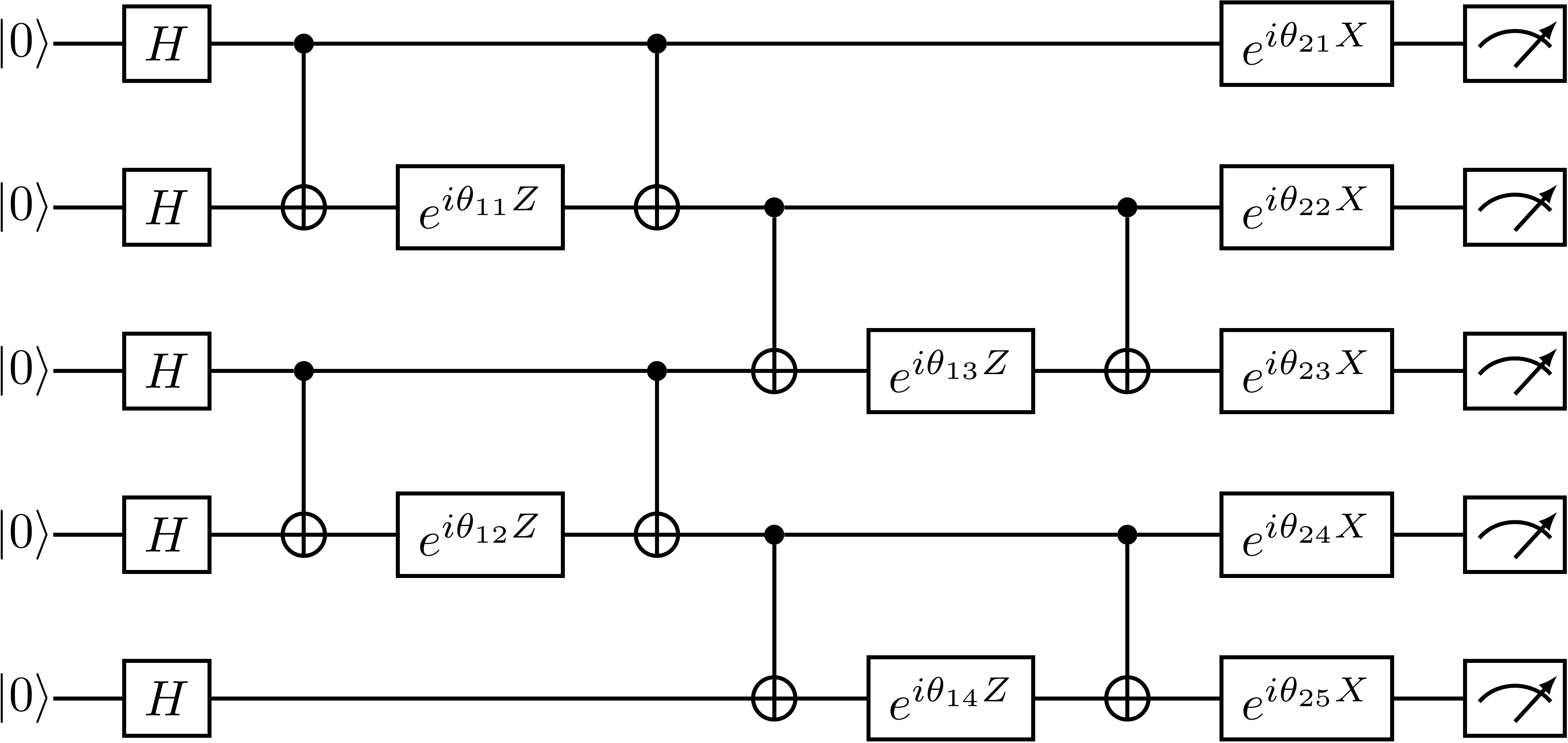}
		\caption{The PQC ansatz utilized in the task to simulate the 5-site TFIM model for both VQE and VQNHE. This quantum circuit is evaluated on noiseless simulator, noisy simulator and quantum hardware IBM\_Santiago.}
		\label{fig:santiago}
	\end{figure}
	
	Since the quantum-classical hybrid optimization process is rather time consuming to run in real quantum hardware, we first optimize both VQE and VQNHE using the ideal simulator without measurement uncertainty and quantum noise to obtain the corresponding optimal parameters for the PQC and the classical model. The ideal converged energies for both VQE and VQNHE are $-5.897229$ (relative error $2*10^{-2}$) and $-6.02667418 $ (relative error $2*10^{-12}$), respectively. If we run the corresponding PQC on noiseless simulator, the results obtained from bitstrings have standard deviations originated from measurement uncertainty while the mean values stay the same as the ideal cases within the error bar for both cases. The number for measurement shots in each group is $8192$: namely, we run 100 groups of independent measurements and report the mean and standard deviation among these groups of measurement results. The mean value and standard deviation pairs for both VQE and VQNHE in noiseless simulator are: $(-5.895, 0.0207)$, $(-6.023, 0.0236)$. Note that the bitstrings measured in the VQE case are ZZZZZ and XXXXX while the bitstrings measured in the VQNHE case are ZZZZZ, XZZZZ, ZXZZZ, ZZXZZ, ZZZXZ, ZZZZX.
	
	Things become more interesting when quantum noise sets in, such as noisy simulators and the real quantum hardware. Note that the noisy simulator we utilized in Qiskit is also characterized by the noise model for IBM\_Santiago.
	In terms of the standard VQE, the simulation and data processing from measured bitstrings are similar as the noiseless case. The only add-on is that we apply measurement error mitigation for all results from noisy simulator and real quantum hardware. This QEM technique is provided by Qiskit: \url{https://qiskit.org/textbook/ch-quantum-hardware/measurement-error-mitigation.html}.  We compare the results from 100 groups of measurements in the simulator case and 10 groups of measurements in the real hardware. Each group contains 8192 measurement shots. The mean and standard deviation pairs from the quantum simulator and hardware cases are $(-5.724, 0.0282)$ and $(-5.690, 0.0228)$, respectively.
	On the other hand, when we run the VQNHE in noisy settings, the result obtained from the ideal optimal post-processing on the obtained measurements is not optimal anymore. This is because the neural network is trained in ideal scenario and doesn't incorporate the appropriate noise model in quantum hardware. Therefore, retraining on the neural network is required based on the results from measurement bitstrings. This retraining is natural and reasonable in terms of VQNHE training, as in the real problem, we can only train the neural network based on measurement results anyway. From another perspective, if retraining on the post-processing module can further lower the energy estimation, such post-processing is similar to a quantum error mitigation (QEM) workflow. VQNHE might have QEM baked in by adjusting the post-processing to alleviate the effect of quantum noise in the PQC. For the noisy simulator, we again collect 100 groups of measurements ($8192$ shots each group) to determine the mean value and standard deviation of energy estimation. The result pairs before and after neural network retraining are $(-5.858, 0.0290) $ and $ (-5.867, 0.0246)$. In terms of hardware experiments, we collect bitstring on each required basis in 10 groups ($8192$ shots each group), and we compute the mean and standard deviation from a list of groups that are randomly selected by combining measurement groups on each basis.  The result pairs before and after neural network retraining are  $(-5.852, 0.0283)$ and $(-5.894, 0.0306)$. The results after retraining are reported in FIG.2 in the main text as the results of the noisy simulator and hardware. As we can see, the energy estimation is indeed lower with network retraining. In some sense, the observation indicates that quantum error is partially reduced by post-processing retraining. Since the energy gain after neural network retraining is larger in the real hardware case,  retraining technique is more useful to suppress quantum error induced by real hardware which is not fully reproduced by noisy simulators.
	
	Some further observations and comments are in order. The results from the noisy simulator and the real quantum hardware match well in both VQE and VQNHE cases.  The standard deviation that characterizes the measurement uncertainty is similar in both cases. VQNHE has measurement uncertainty in the same order as VQE. This fact implies that the required number of measurement shots to reach a given accuracy is similar for VQNHE and VQE. Variational post-processing is potentially good for QEM but the error mitigation capacity is still limited since the energy gain brought by retraining is not enough to recover the ideal result. It deserves further investigation on the interplay between VQNHE and QEM.

	\subsection{Technical details for the simulation of LiH }
	The Hamiltonian of molecules can be described by the many-electrons' Schr\"odinger equation with Born-Oppenheimer approximation. In typical quantum chemistry setups, we first solve the equation with the Hartree-Fock approximation, where the ground state is assumed to be a single Slater determinant. The Hartree-Fock solution gives the so-called molecular orbitals which is the linear combination of atomic orbitals. We then use these molecular orbitals as the basis functions and transform the Schr\"odinger equation into the second quantization form. At this stage, the coefficients in the Hamiltonian are determined by the overlap integral between molecular orbitals. These values are calculated via Psi4 and OpenFermion-Psi4 plugin. Since we use the standard STO-3G atomic orbitals, there are $(1+5)*2=12$ spin orbitals (independent fermions) in total. We further invoke the idea of the complete active space to restrict the freedoms. For the core orbitals 1s of Li, we assume they are always filled. And since the Li-H bond is mainly formed by the hybridization of H(1s) and Li(2s, 2pz), we regard Li(2px, 2py) as virtual orbitals that are never filled. Confining to the complete active space, we are left with only six fermionic degrees of freedom. After building the fermionic Hamiltonian, we further encode the Hamiltonian into qubits.  There are many well-established encoding schemes such as Jordan-Wigner and Bravyi-Kitaev transformation. However, in our case,  there are two conserved quantities, i.e., there are always an odd number of electrons in the spin up and spin down orbitals, we choose to further reduce the Hamiltonian to a 4-qubit representation using a parity conserved binary encoding provided by OpenFermion (See details for binary codes usage in \url{https://quantumai.google/openfermion/tutorials/binary_code_transforms}). The target Hamiltonian contains 100 terms of Pauli strings, and the coefficients before these Pauli strings may change as functions of bond distances.
	
	\begin{figure}[t]\centering
		\includegraphics[width=0.7\textwidth]{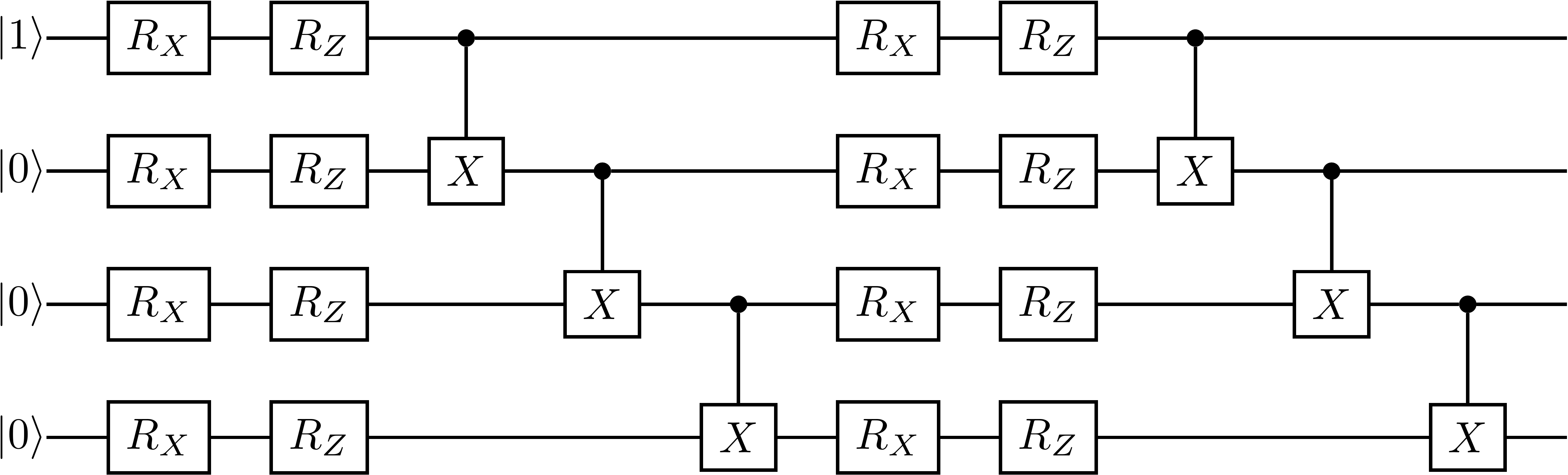}
		\caption{Depth 2 hardware efficient ansatz utilized for VQNHE and VQE optimization on LiH. The input initial state is $\vert 1000\rangle$ so that when the state goes through two rounds of CNOT gates, the output state is approximately $\vert 1010\rangle$ when all rotational angles are small. Such state is the Hartree-Fock mean field solution under our symmetric binary qubit encoding.}
		\label{fig:hea}
	\end{figure}
	
	Next, We optimize the target Hamiltonian energy with VQE and VQNHE. Both methods utilize the PQC as shown in Fig.~\ref{fig:hea}. The parameters $\vect{\theta}$ are initialized near zero, so that the output state from this PQC is in the form $\vert 1010\rangle$, which is the Hartree-Fock solution (two electrons filled in the lowest molecule orbitals within the complete active space) under the binary encoding used in this case. With such an initialization trick and $\vert 1000\rangle$ as the input state, VQE and VQNHE give stable performance and somehow avoid the issue of barren plateaus, since HF solution is generally a very good approximation for such small molecules. Besides, for the VQNHE case, the post-processing model is a lightweight fully connected neural network with two hidden layers of units $8$ and ReLU activation. The final scalar output is activated by $e^{\phi_0\tanh(*)}$, where $\phi_0$ is a trainable weight for $f$ and can be restricted in the norm to avoid large measurement fluctuations as we mentioned in the section above. We run $10\sim 30$ VQE and VQNHE simulations ( with different initializations) for each bond distance, and keep the lowest energy as the final converged energy value. In addition, we utilized the idea of adiabatic VQE, where the initialization from the best result in the last round is perturbed and serves as the initialization for the next Hamiltonian parameter when the bond distance is updated from the last round. This trick boosts the convergence of overall VQE training.
	
	More on hyperparameters in VQNHE for LiH. Default initialization for parameters in the hardware efficient ansatz is drawn from a Gaussian distribution with a standard deviation of 0.2. We also perform initialization from the best initialization of the last and the second last sets of parameters by perturbation with a uniform noise with width 0.1. The cutoff for $\phi_0$ is set to be 5 with an initial value of 1 while, in practice, the converged result often shows much smaller $\phi_0$ indicating smaller fluctuations with measurement outcomes. The energy convergence threshold is set at $10^{-9}$, Adam optimizers are utilized. The learning curves for both quantum and classical modules are shown in Fig.~\ref{fig:lclih}.
	
	\begin{figure}[t]\centering
		\includegraphics[width=0.4\textwidth]{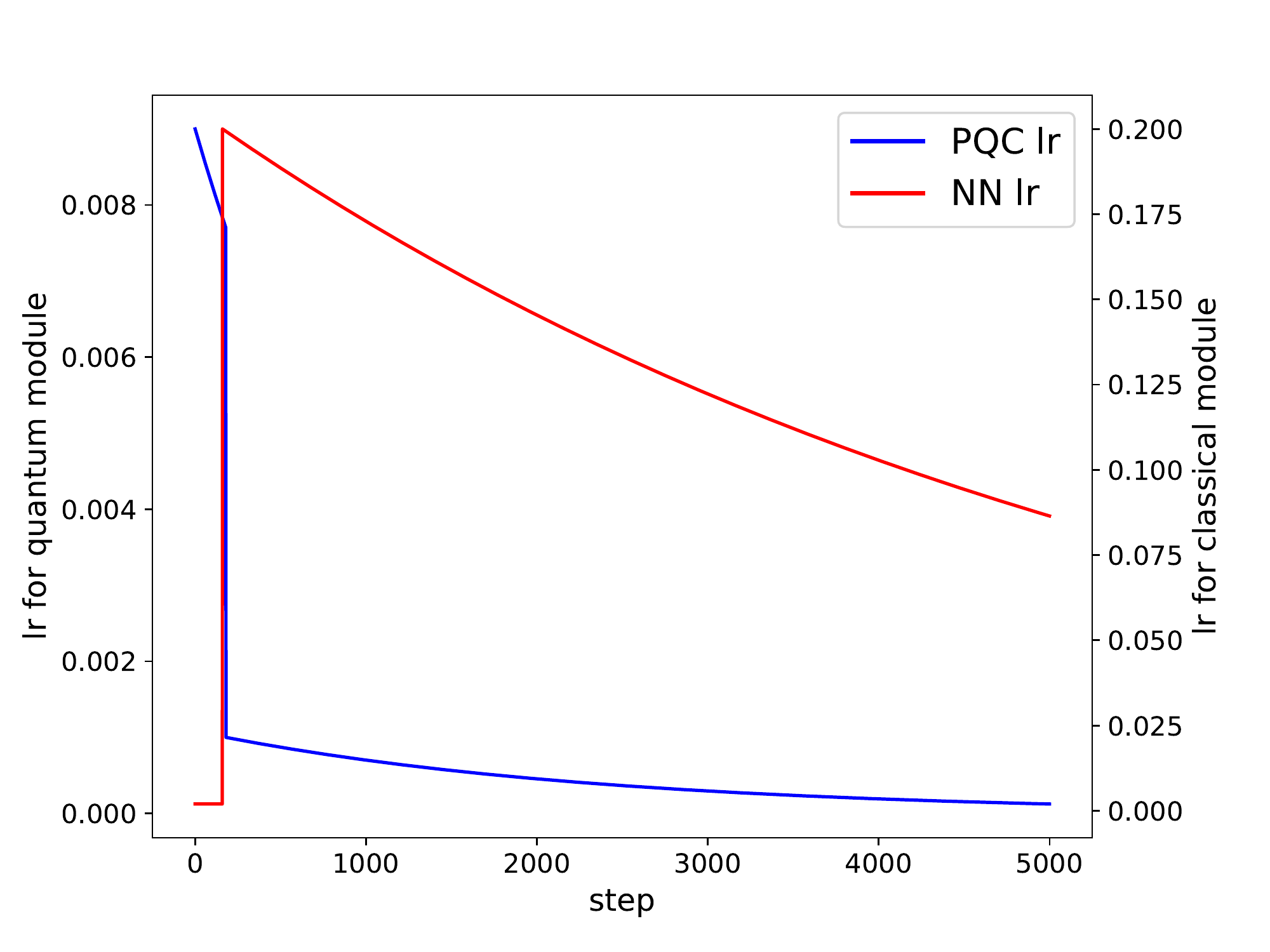}
		\caption{The learning rates of both Adam optimizers for the PQC and neural network (NN) with different update steps in LiH energy optimization. Quantum spin model optimization shares similar learning curves though with slightly different value.}
		\label{fig:lclih}
	\end{figure}

	\subsection{Technical details for the simulation of H$_6$-hexagon and H$_6$-chain }
	
		 \begin{figure}[t]\centering
		\includegraphics[width=0.65\textwidth]{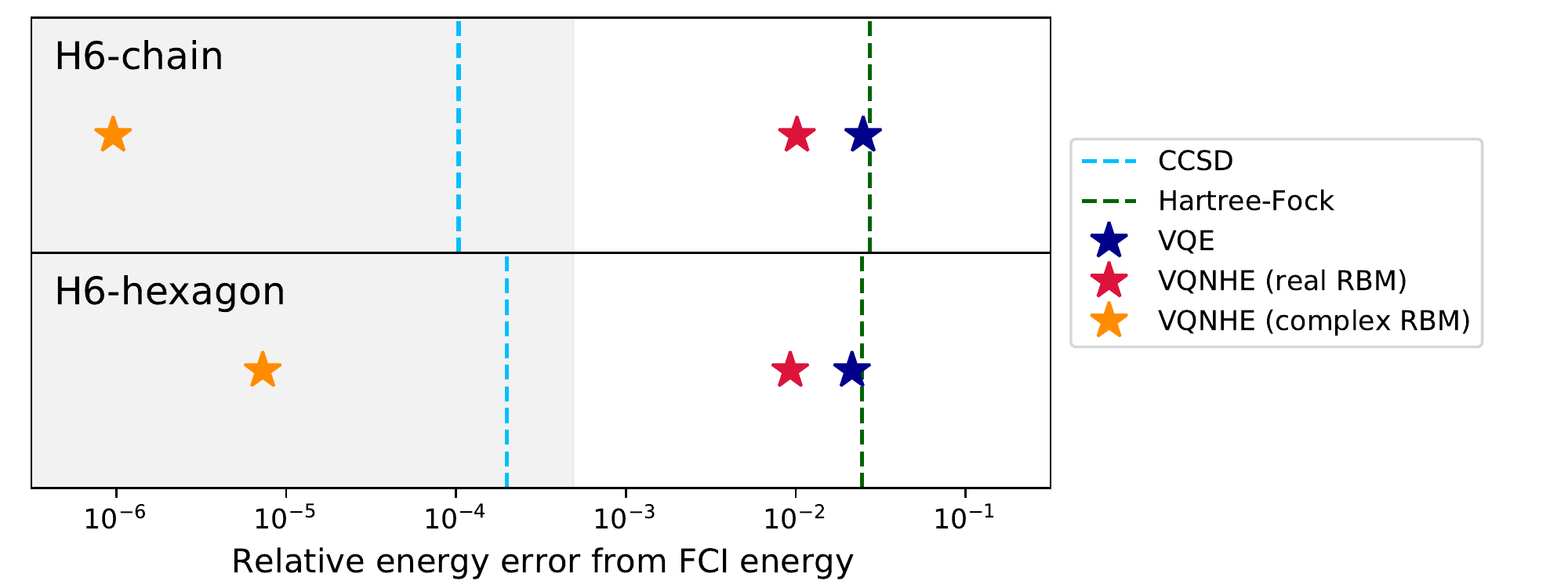}
		\caption{Relative energy errors in H$_6$-chain and H$_6$-hexagon system. Energies obtained by Hartree-Fock method, CCSD method, VQE method, VQNHE method with both real and complex RBM post-processing are shown. The baselines for the relative error are obtained by FCI method. The gray shaded areas are within chemical accuracy. VQNHE with complex RBM post-processing even outperforms CCSD method.}
		\label{fig:h6}
	\end{figure}
	
	We follow similar procedures as in the case of LiH simulation to obtain the qubit Hamiltonian for both H$_6$-hexagon and H$_6$-chain molecules. The bond distances are fixed at 0.99 Å and 0.93 Å, respectively, as these distances should correspond to the equilibrium configurations for these systems. There are 12 spin orbitals for each molecule if the STO-3G basis set is used. Since the systems are not spin polarized, with the symmetry enforced qubit encoding, we can translate the electronic Hamiltonian to a qubit Hamiltonian using only 10 qubits. Namely, we save one qubit for each spin-up and spin-down subspaces for electrons.

	The PQC utilized in both VQE and VQNHE is depth-4 hardware efficient ansatz, consisting of circuit layers that each is composed of Rx layer, Rz layer and CNOT layer. The initial state is given by $\vert 1110100101\rangle$, which is transformed into the Hartree-Fock solution $\vert 1110011100\rangle$ after 4 layers of CNOT gates. Therefore, with such an initial state and small initialization around zero for all rotation gates, the PQC can more easily converge to good approximations.
	As for the classical neural network part for the hybrid representation, we consider both real and complex RBM. The RBM neural architecture is given by:
	\begin{equation}
		f(\vect{s})=e^{\sum_{i} a_{i} s_{i}} \prod_{j=1}^{M} 2 \cosh 
	(	b_{j}+\sum_{i}^{N} W_{i j} s_{i}),
	\end{equation}
where trainable parameters $\vect{\phi} = \{W, a, b\}$. If these parameters are allowed to be complex numbers, we call the module complex RBM. The expressive power of the RBM is determined by the ratio between hidden units and visible units $M/N$. In our setup, $N=10$ and $M=40$.
	It is worth noting that this is the only part where we use complex post-processing modules throughout this work. The reason is that the phase structure of the ground state for such complex systems is hard to capture by the shallow circuit, and a more powerful classical post-processing module may share the burden with a shallow PQC. As clearly implied by the way this hybrid representation of quantum states is setup, as quantum hardware improves we may rely more on the expressive power of a deeper quantum circuit and refrain from using more sophisticated neural networks on the classical end.
	 
	 We run the simulation using an ideal simulator, where more than 700 terms of Pauli strings in the Hamiltonian are compiled as one big matrix and its expectation value is optimized. Such Hamiltonian compiling technique can greatly reduce the required simulation time and resources. The results for both H$_6$ systems are shown in Fig.~\ref{fig:h6}. The relative energy error is defined as $\abs{(E-E_{\text{FCI}})/E_{\text{FCI}}}$. The results obtained with complex RBM post-processing are impressive as they are even much better than the values obtained from CCSD, which is known as the golden standard in quantum chemistry.

	 More on hyperparameters for VQNHE training on the H$_6$ system. Complex RBM is initialized with all weights around zero with a standard deviation $0.005$. Parameters in the hardware efficient PQC are initialized around zero with a standard deviation $0.03$.  Both optimizers for the PQC and the RBM are Adam. The quantum part is mainly optimized in the first 200 rounds with the learning rate 0.01 and after 200 rounds, the learning rate for the i-th round reads $0.002*0.5^{(i-200)/800}$. The learning rate for updating the RBM is $0.006*0.5^{(i-200)/20000}$ after 200 rounds and $0.0006$ for the first 200 optimization rounds.
	 
	 Lastly, we comment that the energy optimization curve in the VQNHE case is very special. The optimization can make no progress for thousands of rounds and, suddenly, the energy will start to decrease again. Therefore, one needs very strict convergence criteria and must be patient during the optimization.

\end{widetext}

\end{document}